\documentclass[12pt]{spieman}  % 12pt font required by SPIE;
\usepackage{amsmath,amsfonts,amssymb}
\usepackage{graphicx}
\usepackage{setspace}
\usepackage{tocloft}
\usepackage{comment}
\usepackage{lineno}
\title{Laboratory demonstration of a Photonic Lantern Nuller in monochromatic and broadband light}

\author[a]{Yinzi Xin} %ORCID 0000-0002-6171-9081
\author[a]{Daniel Echeverri} %ORCID 0000-0002-1583-2040
\author[a]{Nemanja Jovanovic} %ORCID 0000-0001-5213-6207
\author[a,b]{Dimitri Mawet} %ORCID 0000-0002-8895-4735
\author[c]{Sergio Leon-Saval} %ORCID 0000-0002-5606-3874
\author[d]{Rodrigo Amezcua-Correa}
\author[d]{Stephanos Yerolatsitis}
\author[e]{Michael P. Fitzgerald} %ORCID 0000-0002-0176-8973
\author[a]{Pradip Gatkine} %ORCID 0000-0002-1955-2230
\author[e]{Yoo Jung Kim} %0000-0003-1392-0845
\author[e]{Jonathan Lin} %ORCID 0000-0001-8542-3317
\author[f]{Barnaby Norris} %ORCID 0000-0002-8352-7515
\author[b]{Garreth Ruane} %ORCID 0000-0003-4769-1665
\author[g]{Steph Sallum}

\affil[a]{Department of Astronomy, California Institute of Technology, 1200 E California Blvd, Pasadena, CA, 91125, USA}
\affil[b]{Jet Propulsion Laboratory, California Institute of Technology, 4800 Oak Grove Drive, Pasadena, CA, 91109, USA}
\affil[c]{Sydney Astrophotonic Instrumentation Laboratory, School of Physics, The University of Sydney, Sydney, NSW 2006, Australia}
\affil[d]{The College of Optics and Photonics, University of Central Florida, 4304 Scorpius St, Orlando, FL 32816}
\affil[e]{Department of Physics \& Astronomy, 430 Portola Plaza, University of California, Los Angeles, CA 90095, USA}
\affil[f]{Sydney Institute for Astronomy, School of Physics, Physics Road, The University of Sydney, NSW 2006, Australia}
\affil[g]{Department of Physics \& Astronomy, University of California, Irvine, 4129 Frederick Reines Hall, Irvine, CA 92697 USA}

\cftpagenumbersoff{figure}
\cftpagenumbersoff{table} 
\begin{document} 
\maketitle

\begin{abstract}
Photonic lantern nulling (PLN) is a method for enabling the detection and characterization of close-in exoplanets by exploiting the symmetries of the ports of a mode-selective photonic lantern (MSPL) to cancel out starlight. A six-port MSPL provides four ports where on-axis starlight is suppressed, while off-axis planet light is coupled with efficiencies that vary as a function of the planet's spatial position. We characterize the properties of a six-port MSPL in the laboratory and perform the first testbed demonstration of the PLN in monochromatic light (1569 nm) and in broadband light (1450 nm to 1625 nm), each using two orthogonal polarizations. We compare the measured spatial throughput maps with those predicted by simulations using the lantern’s modes. We find that the morphologies of the measured throughput maps are reproduced by the simulations, though the real lantern is lossy and has lower throughputs overall. The measured ratios of on-axis stellar leakage to peak off-axis throughput are around $10^{-2}$, likely limited by testbed wavefront errors. These null-depths are already sufficient for observing young gas giants at the diffraction limit using ground-based observatories. Future work includes using wavefront control to further improve the nulls, as well as testing and validating the PLN on-sky.
\end{abstract}


% Include email contact information for corresponding author
{\noindent \footnotesize\textbf{*}Yinzi Xin,  \linkable{yxin@caltech.edu} }

\begin{spacing}{2}   % use double spacing for rest of manuscript

\section{Introduction}
\label{sect:intro}
The characterization of exoplanets was identified by the Decadal Survey for Astronomy and Astrophysics 2020 as one of the top scientific priorities \cite{NRC_2020Decadal}. High-resolution spectroscopy is especially critical for many measurements, including that of the planet's radial velocity, spin, atmospheric composition, and surface features through Doppler imaging \cite{wang_hdc1}. It can also enable the potential detection of exomoons \cite{ruffio_exomoons}. The Photonic Lantern Nuller \cite{xin_2022, Tuthill2022-NIH, xin_spie_2023} is an instrument concept that enables the characterization of exoplanets at and within $1 \,\ \lambda/D$, where $\lambda$ is the wavelength and $D$ the telescope diameter. It is inspired by the Vortex Fiber Nuller (VFN) \cite{Ruane2018_VFN,echeverri_2019}, but unlike the VFN, which has only one nulled channel with a circularly symmetric coupling profile, the PLN provides four nulled channels, each with a unique coupling profile. This allows for more planet flux to be retained, and also allows for constraints to be placed on the planet's flux ratio and spatial position \cite{xin_2022}. Like the VFN, the PLN's ports can also be routed to a high-resolution spectrograph to spectrally characterize the exoplanet.

The PLN exploits the symmetries of the ports of a mode-selective photonic lantern (MSPL) \cite{LeonSaval_MSPL, Velazquez-Benitez:15}, a special type of photonic lantern \cite{LeonSaval_PL_2013, fontaine_2022} that utilizes dissimilar cores, enabling the ports to be mapped into linearly polarized (LP) modes, or the eigenmodes of a radially symmetric, weakly guiding step-index waveguide. Each mode at the multi-mode (MM) face of the lantern is mapped to a single-mode fiber (SMF) output, such that light coupling to a given mode at the MM face side will result in flux in the corresponding SMF core. The device is bi-directional, so light injected into one of the SMF ports will propagate into the mode corresponding to that port at the MM face.

The operating principles of the PLN are fully derived in Ref. \citenum{xin_2022}, where we show that while MSPLs with different port numbers exist, a MSPL with six ports (corresponding to the first six LP modes) provides a good balance of manufacturability and planet throughput, and adding ports beyond the first six does not meaningfully improve the planet throughput. Here, we summarize the case where a telescope beam is injected directly into the MM face end of the photonic lantern (without any optical vortex in the beam). If we label the LP mode azimuthal order by an integer $m$ analogously to how Zernike polynomials are labeled --- i.e. positive $m$ indicating an azimuthal component of $\cos(m\theta)$ and negative $m$ indicating $\sin(m\theta)$ --- then, the polar component of the integral that describes the coupling of an unaberrated on-axis telescope beam into a photonic lantern port is given by

\begin{equation} \label{eq:lp_mode_overlap}
\begin{split}
    \int_0^{2\pi} \cos(m\theta) d\theta, \quad m & \geq 0, \quad \mathrm{or}\\
    \int_0^{2\pi} \sin(m\theta) d\theta, \quad m & < 0.
\end{split}
\end{equation}

This polar integral will integrate to 0 and result in an on-axis null \textit{except} when $\pm m=0$. For a six-port MSPL, the LP 01 and LP 02 ports (which have $m=0$) are non-nulled. The LP 11a ($m=1$) and LP 11b ($m=-1$) ports have a first-order null, and the LP 21a ($m=2$) and LP 21b ($m=-2$) ports have a second-order null. As explored in Ref. \citenum{xin_2022} (and by analogy to the charge 1 and charge 2 VFNs in Ref. \citenum{ruane_2019_spie}), the LP 11 ports are more sensitive to close-in planet signals but also more sensitive to tip-tilt errors, while the LP 21 ports are less sensitive to both. The PLN thus provides measurements with both null orders on the same instrument.

The simulated spatial throughput maps (the fraction of incident light that gets coupled and propagated through each port, as a function of the spatial position of the source) for an ideal six-port MSPL is shown in Figure 1a. For a lossless lantern, the throughput for each port is the same as the fractional coupling into the port. The central nulls (the location where on-axis starlight is canceled) can be seen at the center of each map. Additionally, as seen in the x-axis cross-section shown in Figure 1b, the summed throughput from all four nulled ports given an off-axis planet at $1.0 \,\ \lambda/D$ is $\sim 60\%$.

In Ref. \citenum{xin_2022}, we also explore adding a vortex mask ahead of the photonic lantern, which results in a different set of ports being nulled. The PLN with a vortex may provide higher detection capabilities depending on the position of the planet and the distribution of wavefront error, but it is more complicated, as the vortex introduces a potential source of misalignment as well as chromaticity \cite{echeverri_2020_spie}. Therefore, in this work, we test the simplest PLN architecture, which does not use a vortex.

\begin{figure*}[t]
\begin{center}
    a \includegraphics[scale = 0.48]{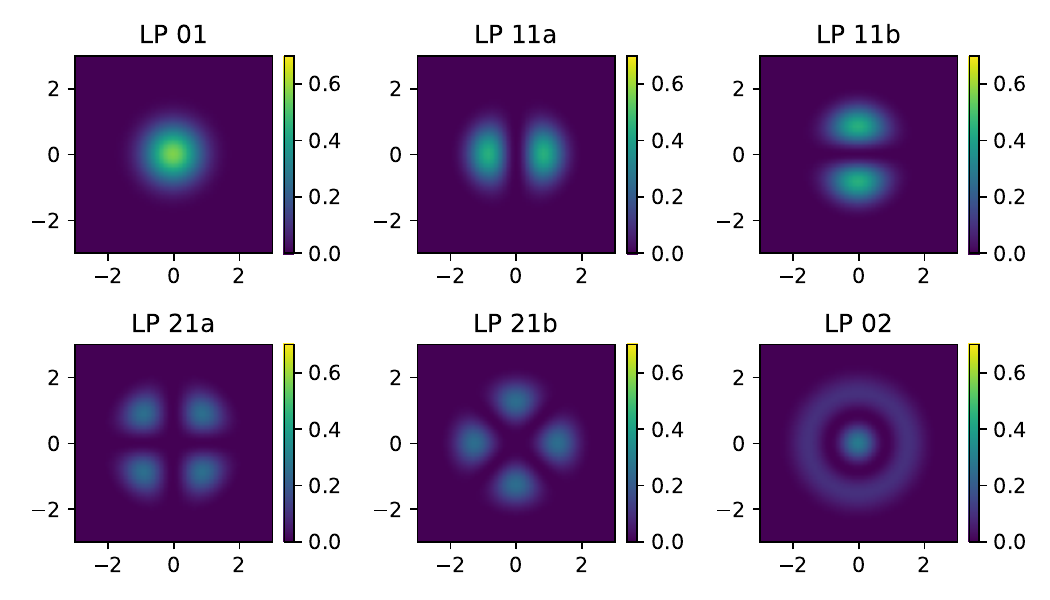}
    b \includegraphics[scale = 0.45]{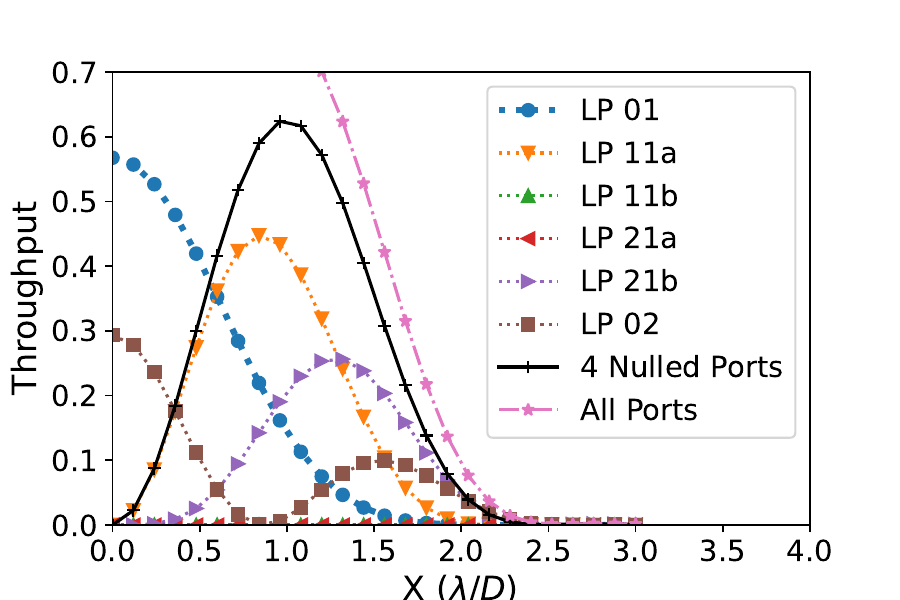}
	\caption{\label{fig:sim_ideal} a) Throughput maps for each port of an ideal six-port mode-selective photonic lantern, spanning from -3 $\lambda/D$ to 3 $\lambda/D$ in each direction. b) X-axis line profiles from the throughput maps. The four nulled ports are LP 11ab and LP 21ab. The line profile of the summed throughput of the nulled ports is shown in black with cross-hair markers. Figure adapted from Ref. \citenum{xin_2022}.
	}
\end{center}
\end{figure*}

Previous conceptual work on the PLN assumed a perfect, lossless MSPL \cite{xin_2022}. However, a real MSPL will not have the ideal mode shapes corresponding to perfect LP modes, and the device itself will have some additional throughput loss. In this work, we characterize the properties of a real photonic lantern including manufacturing imperfections (described in Section \ref{sec:lant_charac}). We then integrate it into a testbed to demonstrate the PLN in the lab (described in Section \ref{sec:pln_demo}).

\section{Lantern Characterization} \label{sec:lant_charac}

A picture of the MSPL (optimized to have six ports at 1550 nm) is shown in Fig. 2a. The lantern is the stiff silver portion in the top right, with the MM end facing towards the right. Each SMF output of the lantern is connected to one of the white fiber pigtails.

\begin{figure*}[t]
\begin{center}
    \includegraphics[scale = 0.45]{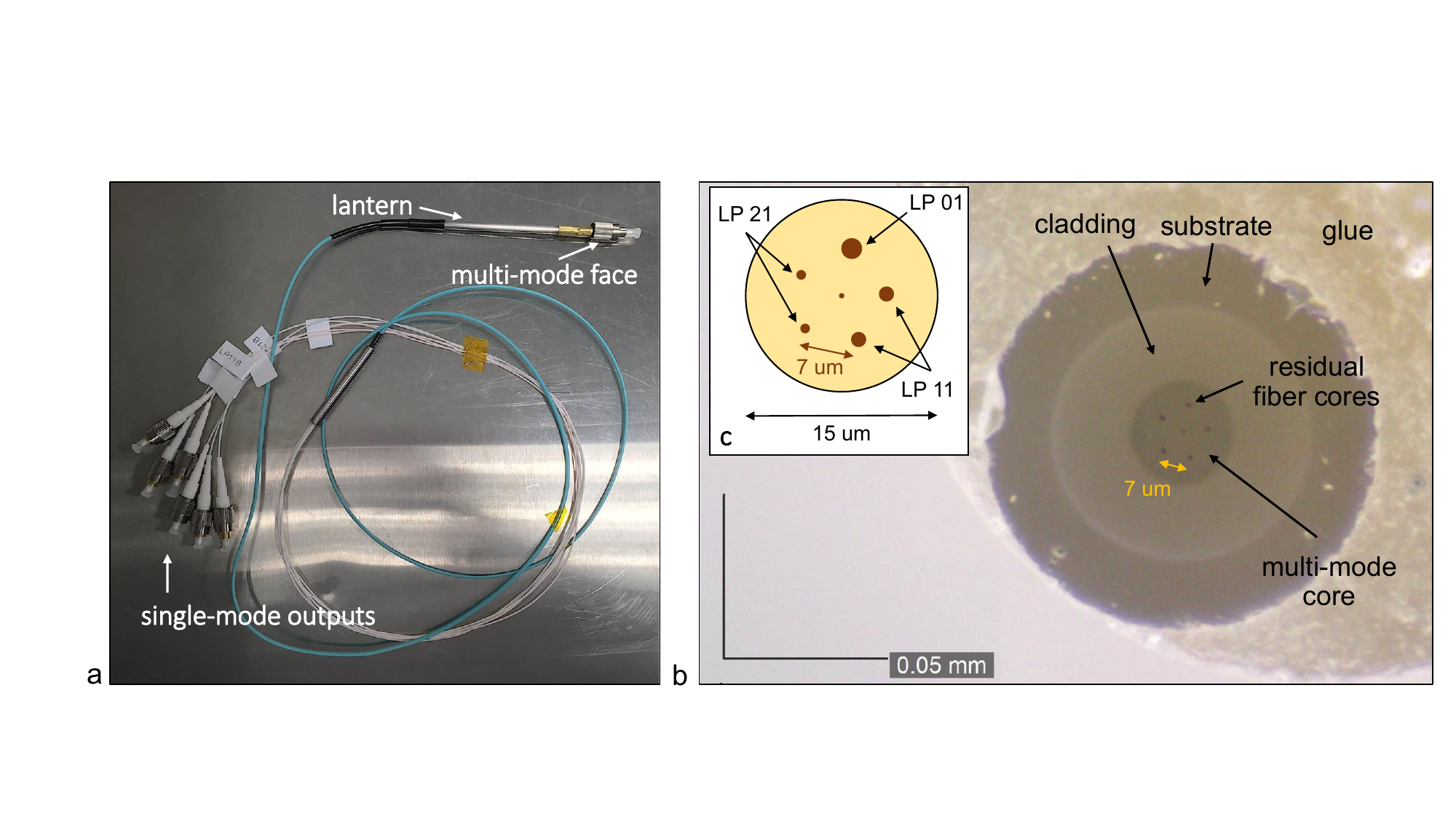}
	\caption{\label{fig:lant_pictures} a) A picture of a six-port MSPL. The lantern is the stiff silver portion in the top right, with the MM end facing towards the right. Each SMF output of the lantern is connected to one of the white fiber pigtails. b) A microscope image of the MM face taken with a Dino-Lite Edge 3.0. The residual fiber cores are arranged in a pentagonal pattern, visible towards the bottom left as small black circles. The residual fiber cores are embedded in the multi-mode core (dark brown), which is surrounded by fiber-doped glass cladding (light brown ring). Surrounding the cladding is a silica substrate (outermost dark brown ring), followed by the glue (rough tan material) that attaches the lantern to the connector. c) A schematic of the design of the MM face for comparison. The observed MM core diameter and distance between adjacent residual SMF cores are both consistent with the design values of 15 $\mu$m and $7 \,\ \mu$m respectively.
	}
\end{center}
\end{figure*}

We first characterize the properties of this lantern on its own by taking microscope images of the MM interface, measuring the throughput through each port, and using an interferometer to reconstruct the mode shapes corresponding to each port. 

\subsection{Microscope Imaging of Lantern Interface}

We use a Dino-Lite Edge 3.0 microscope to image the multimode interface of the lantern, which is shown in Fig. 2b. The residual fiber cores are arranged in a pentagonal pattern, visible towards the bottom left as small black circles. The residual fiber cores are embedded in the multi-mode core (dark brown), which is surrounded by fiber-doped glass cladding (light brown ring). Surrounding the cladding is a silica substrate (outermost dark brown ring), followed by the glue (rough tan material) that attaches the lantern to the connector. The microscope image verifies the expected pentagonal arrangement of the single cores for the MSPL design presented in Ref. \citenum{LeonSaval_MSPL}, also depicted in schematic form in Fig. 2c. The observed MM core diameter and distance between adjacent residual SMF cores are both consistent with the design values of 15 $\mu$m and $7 \,\ \mu$m respectively.

\subsection{End-to-end Throughput Measurements} \label{sec:thpts}

Next, we measured the throughput of each of the lantern's ports from the single-mode inputs to the MM face using a power meter (Thorlabs S122C), with a  laser diode (Thorlabs KLS1550) as the light source. We first took a background measurement with the light turned off. Next, we took a measurement of the power coming out of the fiber directly connected to the laser. Then, we connected the light source fiber to one of the SMF pigtails of the photonic lantern and measured the power coming out of the MM face. After background subtracting the measurements, we take the ratio of the power coming out of the MM face to the power coming out of the source fiber to be the throughput of that port. We repeat this process five times for each port, and report the mean and standard deviation of those five measurements in Table \ref{tab:throughputs}. These throughput measurements are of the entire lantern assembly and include any connectorization losses (e.g. at the interfaces to the LC connectors), splice losses (e.g. between the lantern and the SMF pigtails), and losses from Fresnel reflection and propagation through the pigtails — and are consistent with losses expected from the assembly manufacturing. Note that in comparison, a typical SMF patch cable of equivalent length would have throughput $>95\%$.

\begin{table}
\caption{The average and standard deviation of five throughput measurements for each lantern port. Note that these throughput measurements are of the entire lantern assembly, including any connectorization losses or losses through the SMF pigtails, such as Fresnel loss and propagation loss.}
\begin{center} \label{tab:throughputs}
\begin{tabular}{|l|c|c|}
 \hline
 Port \hspace{1cm} & Throughput & Standard Deviation \\ 
 \hline
 LP 01  & 0.886 & 0.013 \\ 
 LP 11a  & 0.890 & 0.013 \\ 
 LP 11b  & 0.778 & 0.017 \\ 
 LP 21a  & 0.583 & 0.009 \\ 
 LP 21b  & 0.614 & 0.022 \\ 
 LP 02  & 0.617 & 0.014 \\ 
 \hline
\end{tabular}
\end{center}
\end{table}

\subsection{Characterization of Modes} \label{sec:oah}

We use a technique called off-axis holography (OAH) to measure the complex electric field corresponding to each of the lantern's ports. A detailed discussion of the principles of OAH can be found in Ref. \citenum{cuche_00}. In summary, a broad reference beam is interfered with an image of the lantern mode, creating fringes across the image. The fringes create sidelobes in Fourier space, and by filtering the Fourier-transformed signal, the electric field of the mode can be reconstructed. The principles and process of OAH is also similar to that of the self-coherent camera \cite{baudoz_2006_scc}.

A picture of our optical setup for OAH is shown in Fig. 3. For this experiment, we use a polarized tunable narrow linewidth laser (Thorlabs TLX2), set to a wavelength of 1568.772 nm with a linewidth of 10 kHz, which equates to $\sim 10^{-16}$ m. This results in a coherence length of $\sim 10^4$ m. The light is split by a 50:50 polarization-maintaining (PM) splitter, which sends half the light to the imaging arm of the interferometer, and the other half to the reference beam arm. The light in the imaging arm goes through a polarization controller, then to one of the single-mode inputs of the MSPL. The light coming out of the MM face of the lantern goes through a lens that collimates the beam, then to a lens that focuses the image onto an InGaAs camera (First Light C-RED 2), which has a pixel pitch of 15 $\mu$m.

The light in the reference beam path passes through a PM fiber coil (to better match path lengths between the two arms, even though the long coherence length does not necessitate this), then through a lens that forms a diverging beam large enough to cover the entire lantern mode image. This reference beam interferes with the lantern mode image, creating fringes that allow us to retrieve the complex mode field using Fourier analysis. Ideally, the reference beam should be collimated into a flat wavefront; however, we have simply made it large enough that its phase is slowly varying over the extent of the mode image, so does not significantly impact the reconstructions.

The visibility of the fringes is highest when the polarization state of the two beams are matched. To match the polarization between the two arms, a calibration polarizer is first inserted into the reference arm. The polarizer is then fixed to the angle that cancels the flux of the reference beam on the detector. Then, the polarizer is moved into the lantern arm, and the polarization controller set to minimize the flux that goes through the polarizer and onto the detector. This process aligns the polarization states of both arms to each other. The calibration polarizer is then removed from the beam for the rest of the experiment.

\begin{figure*}[t]
\begin{center}
    \includegraphics[scale = 0.58]{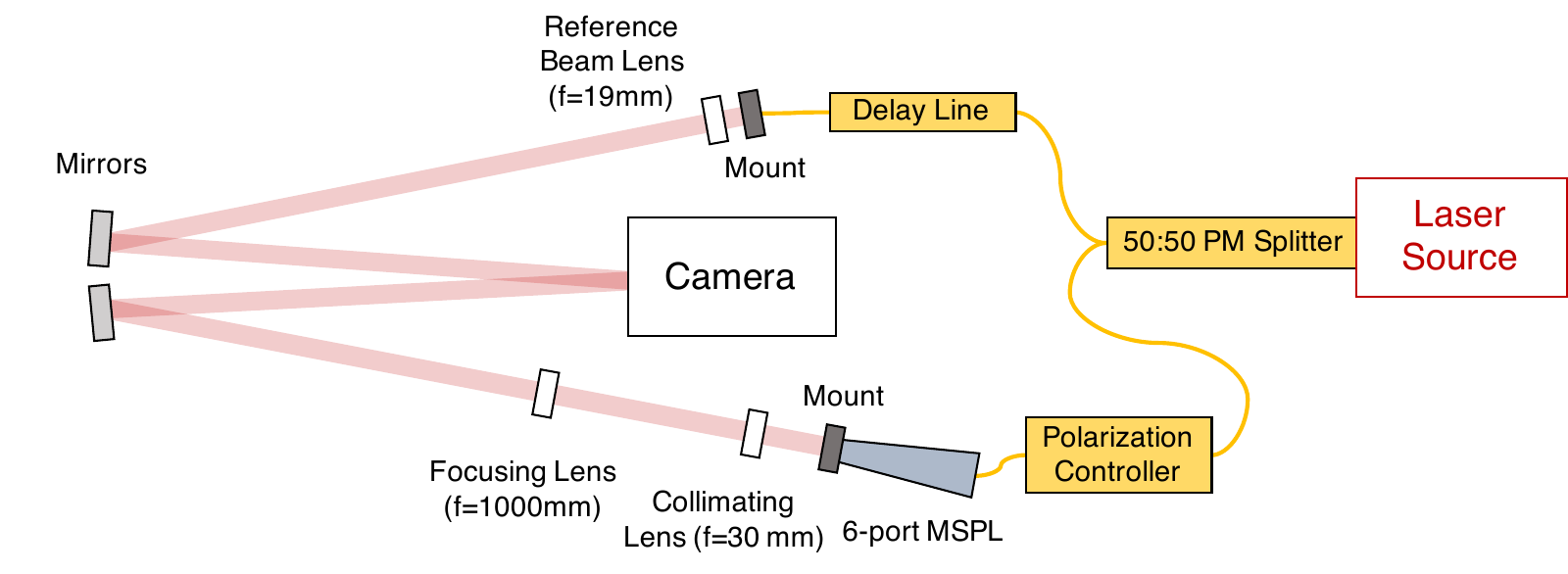}
	\caption{\label{fig:oah_setup} The optical setup for OAH measurements of the lantern modes. The light from the laser is split by a 50:50 polarization-maintaining splitter. Half the light is sent to the imaging arm, through a polarization controller, then to one of the single-mode inputs of the MSPL. The light coming out of the MM face  of the lantern goes through a lens that collimates the beam, then to a lens that focuses the image onto an infrared camera. The other half of the light is sent to the reference beam path, where it passes through a PM fiber coil delay line, then through a lens that creates a diverging beam large enough to cover the entire lantern mode image. The polarization controller is used to match the polarizations of the two beams. The reference beam interferes with the lantern mode at the detector, creating fringes that allow us to retrieve the complex mode field using Fourier analysis.
	}
\end{center}
\end{figure*}

See Ref. \citenum{cuche_00} for the principles of using Fourier analysis to retrieve complex amplitudes from off-axis holography (OAH). For our work, an example hologram of the LP 21b port (after dark subtraction and centering) is shown in Fig. 4a, and the same hologram zoomed in to the center (such that the fringes are visible) is shown in Fig. 4b. Note that the fringes of interest resulting from interference between the two beams are the fine horizontal ones. The faint, wide, vertical stripes are not from the interference of the two beams, but rather the structure of the reference beam itself, as shown in Fig. 4c. We apply a 2D Fourier transform to the (full-sized) hologram to obtain the Fourier space signal (Fouriergram) shown in Fig. 4d. We then isolate the top right lobe of the Fouriergram, shown in Fig. 4e.

The cropped lobe in Fig. 4e is centered to minimize the amount of tip-tilt signal in the final reconstructed mode, shown in Fig. 4f. To obtain this final reconstructed mode, we first apply a broad Gaussian window (with a $\sigma$ of 27 pixels) to the cropped Fouriergram to filter out edge effects, then apply an inverse Fourier transform. We then divide by the square-root of the reference beam intensity to remove the impact of its non-uniformity. Finally, we normalize the mode such that its summed intensity is 1. Because the arms of the interferometer are long, and the path length difference between the reference arm and the lantern arm fluctuates widely (due to vibrations and other bench instabilities), we are unable to constrain the global phase of the modes (i.e, the uniform phase term, or the phase piston as expressed in the focal plane). Fortunately, we do not need to know the global phase to predict the coupled intensities, since they are not impacted by the global phases of each mode.

\begin{figure*}[t]
\begin{center}
    \includegraphics[scale = 0.9]{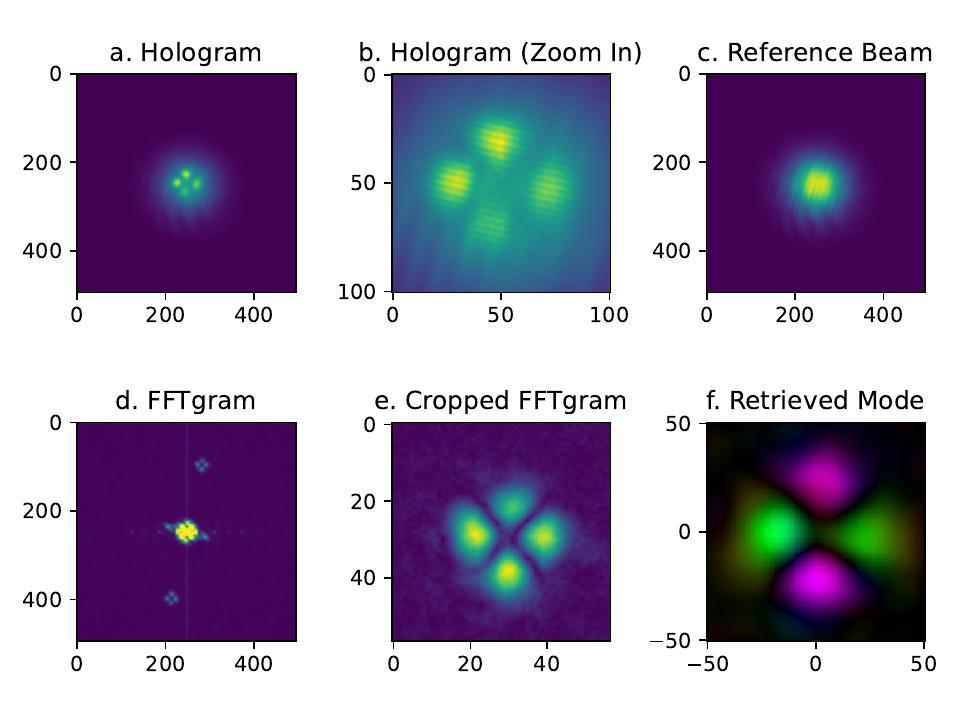}
	\caption{\label{fig:oah_process} a) The measured hologram of the LP 21b port, centered and background subtracted. b) The same hologram as part (a), but zoomed in to the center to show the fringes. Note that the fringes of interest resulting from interference between the two beams are the fine horizontal ones. c) The reference beam intensity, plotted on the same spatial scale as part (a), showing that the faint, wide, vertical stripes are not from the interference of the two beams, but rather the structure of the reference beam itself. d) The 2D Fourier transform of the hologram in part (a). e) The Fourier-space signal in part (d), cropped to the top right lobe. This lobe is centered to minimize the tip-tilt signal in the final reconstructed mode. f) The final reconstructed mode, obtained by first applying a Gaussian window with a $\sigma$ of 27 pixels to part (d) to filter out edge effects, then a 2D inverse Fourier transform. The signal is then divided by the square root of the reference beam intensity to remove its impact, then normalized to a total intensity of 1. The amplitude is indicated by brightness, and the phase indicated by hue.
	}
\end{center}
\end{figure*}

We apply this same reconstruction process to the other 5 modes of the lantern. We also obtain and analyze three separate datasets taken across multiple days and confirm that the measurements are qualitatively stable. After matching the global phases between the three different measurements, we take their mean and re-normalize each mode to a total intensity of 1 to obtain the final reconstructions, shown in Figure 5a. In Figure 5b, we plot the dot-products between the measured modes, which show that they are orthogonal as predicted (the median of the dot-product magnitudes between two different modes is 0.011, which is commensurate with our measurement uncertainty for the mode shapes themselves).

\begin{figure*}[t]
\begin{center}
    \includegraphics[scale = 0.48]{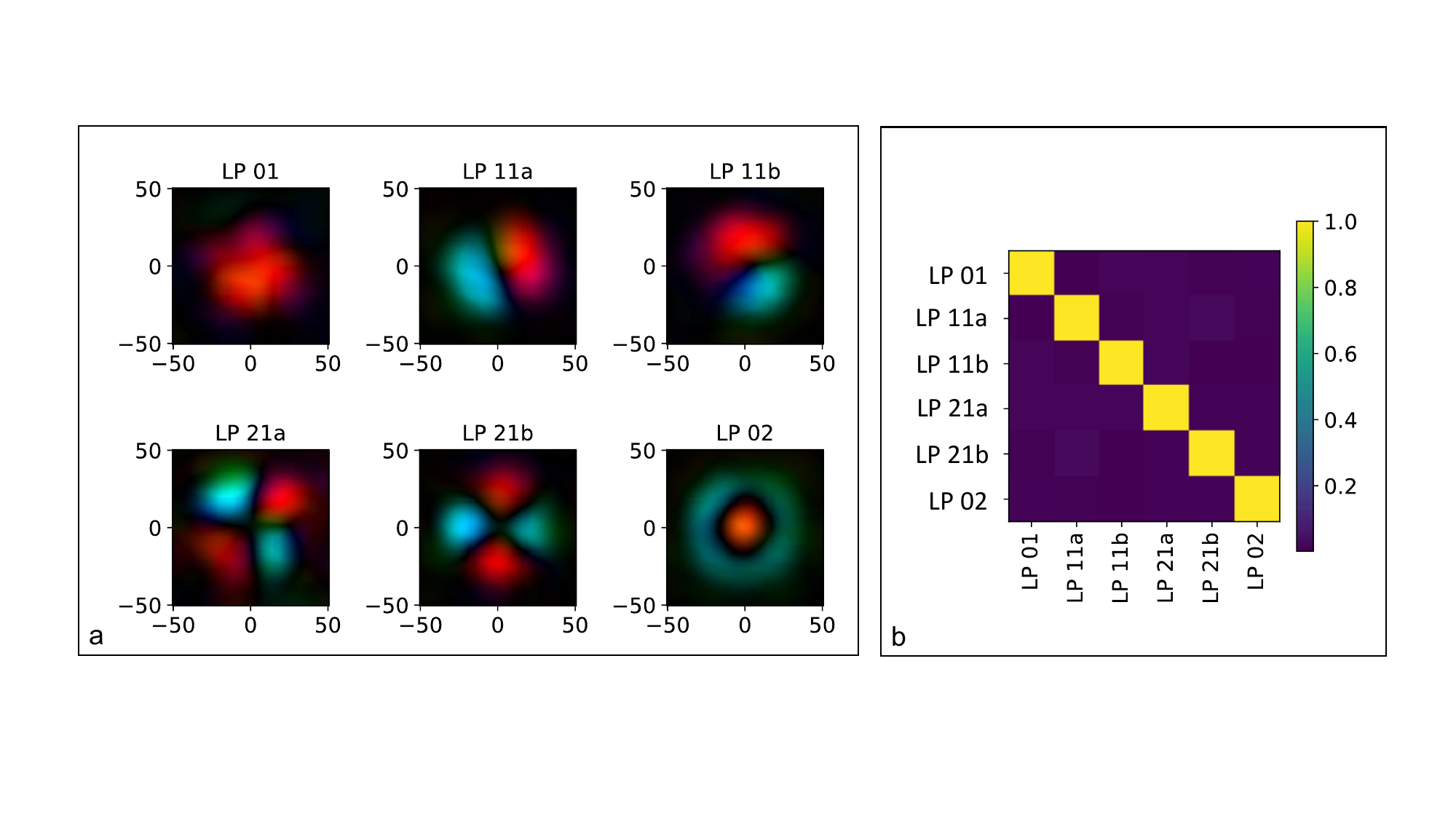}
	\caption{\label{fig:oah_results} a) The measured modes corresponding to each port of the MSPL, obtained using OAH. The amplitude of each mode is indicated by brightness, and the phase indicated by hue. Each mode has its global piston phase term removed, and has been normalized to a total intensity of one. The axes correspond to pixels on the camera, centered about zero. b) The dot-products between the measured modes, which show that they are orthogonal as predicted. The median of the dot-product magnitudes between two different modes is 0.011, which is commensurate with our measurement uncertainty for the mode shapes themselves.
	}
\end{center}
\end{figure*}

In Section \ref{sec:pln_results}, the throughput maps simulated using these measured modes are compared to the actual throughput maps obtained on the testbed.

\section{Photonic Lantern Nuller Demonstration} \label{sec:pln_demo}

After characterizing the properties of the MSPL, we integrated it into the Polychromatic Reflective Testbed (PoRT) \cite{echeverri_2020_spie} at Caltech to demonstrate using it as a nuller.

\subsection{Experimental Setup}

A diagram of the PoRT testbed is shown in Fig. 6. A light source is fed into the bench with a single-mode fiber mounted to the source stage. The light is collimated by an off-axis parabola (OAP) mirror. The collimated light is filtered by a baffle before reflecting off of a $12\times12$ Boston Micromachines deformable mirror (DM). Then, a set of relay OAPs magnifies the beam. In the resulting collimated beam is a mask mount that can be used to insert a pupil plane mask (such as a vortex); however, we leave it empty for this work. The beam then passes through an adjustable-size iris, which we use to control the $F\#$ of the system. The iris aperture diameter ($D$) can range from 1-15 mm, which, given the injection focal length ($f$) of 54.4 mm, can provide focal numbers ($F\# = f/D$) ranging from approximately 3.6 to 55.

The beam is then focused by the last OAP onto the injection stage, which holds both a SMF and the 6-port MSPL. The injection stage can move in translation in all three axes, which allows light to be injected into either the SMF or the MSPL, and can also be used to scan the face of either optic (the fiber or the lantern) across the focused beam. To measure the coupled flux through the SMF, the output end of the SMF is routed to an InGaAs (Femto OE-200-IN2) photodiode. To measure the coupled flux through one of the lantern ports, the corresponding SMF pigtail is routed to the Femto photodiode. This setup can only measure one port at a time, so the measurements for the different ports are made sequentially.

To normalize the data, we use a retractable stage to insert a power meter (Thorlabs S122C) into the beam just before the injection mount to measure the incident flux. After calibrating the readings to that of the Femto photodiode, these beam flux measurements can be used to normalize the coupled flux measurements, in order to obtain throughput measurements.

\begin{figure*}[t]
\begin{center}
    \includegraphics[scale = 0.45]{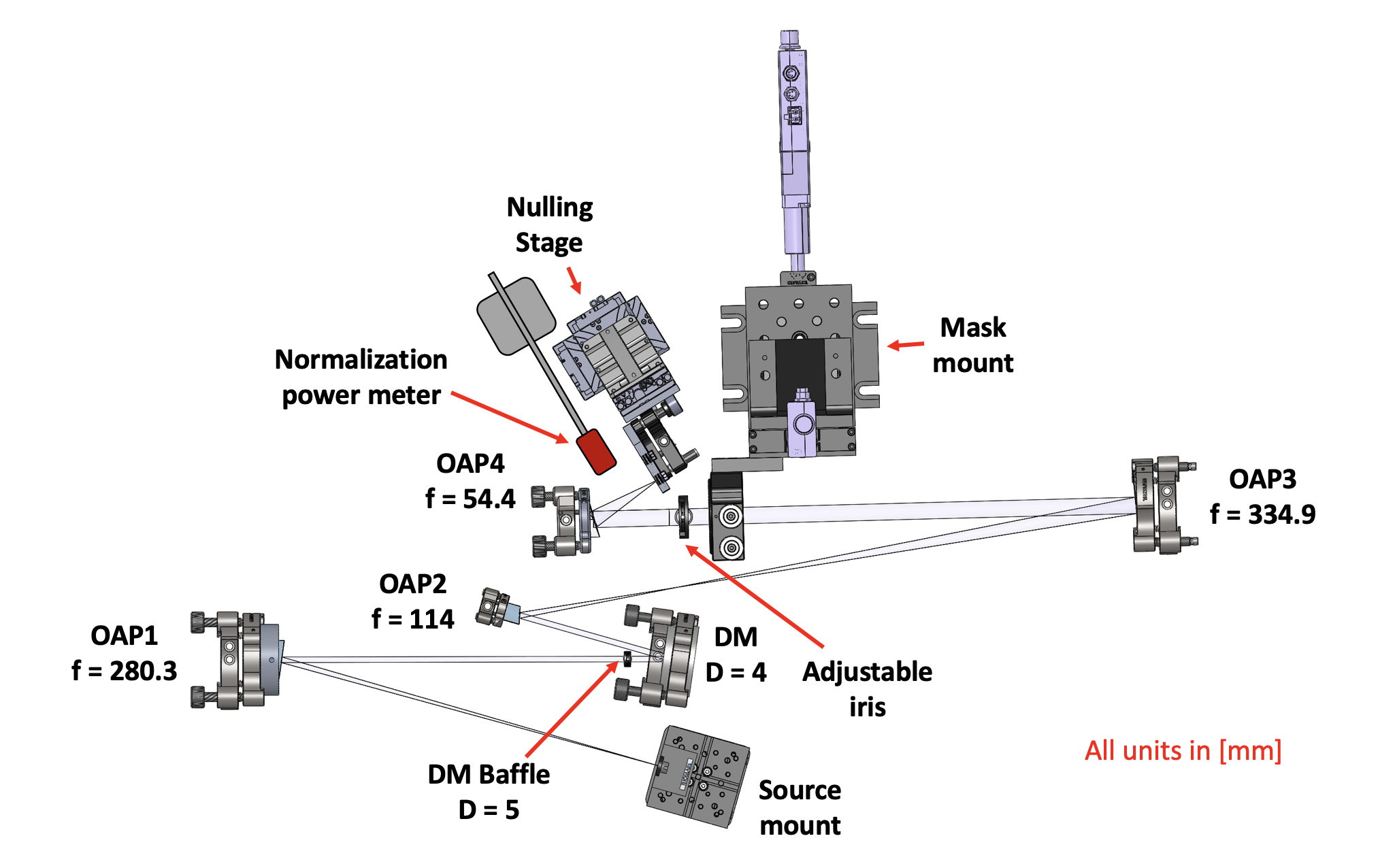}
	\caption{\label{fig:port_diagram} A light source is fed into the bench with a SMF mounted to the source stage. The light is collimated by an off-axis parabola (OAP) mirror. The collimated light is filtered by a baffle before reflecting off of a $12\times12$ Boston Micromachines deformable mirror (DM). Then, a set of relay OAPs magnifies the beam. We leave the mask mount empty for this work. The beam then passes through an adjustable-size iris, which we use to control the $F\#$ of the system. The beam is then focused by the last OAP onto the injection stage, which holds both a single-mode fiber and the 6-port MSPL. A Femto OE-200 photodiode is used to measure the coupled flux. A Thorlabs S122C power meter on a retractable stage can be inserted into the beam just before the injection mount to measure the incident flux, which can be used to normalize the coupled flux measurements for throughput measurements.
	}
\end{center}
\end{figure*}

To best compare our results to the predictions from simulation, we wish to inject a beam with a flat wavefront into the photonic lantern. To flatten the wavefront, we use the SMF as a wavefront calibrator, since coupling into an SMF is maximized when the wavefront is flat. This is most easily seen by expressing the coupling as proportional to the square of the overlap integral of the electric field with the SMF mode (approximated as a Gaussian), expressed in the pupil plane with radial coordinate $\rho$ and angular coordinate $\phi$:

\begin{equation}
\eta \propto \bigg| \int_0^D e^{- \left( \frac{\pi D_f \rho}{2} \right)^2} e^{i \Phi(\rho,\phi)} dA \bigg|^2.
\end{equation}

Here, $D$ is the diameter of the aperture (assumed to be circular), $D_f$ is the mode-field diameter of the SMF, and $\Phi$ is the wavefront error. The coupling is thus maximized when $\Phi$ is uniform over the entire aperture (i.e. there is no wavefront error), as any deviations from uniformity will reduce the value of the integral.

We first optimize coupling into the SMF by adjusting the X, Y, and Z directions of the injection stage. We then optimize the iris size, achieving maximum injection with an $F\#$ of 3.8. Then, we tune the twelve lowest-order Zernike modes of the DM map. We obtain a peak throughput of 74.1\%, whereas the theoretical coupling (calculated as the overlap integral of the ideal SMF mode with the focal-plane electric field given a perfectly flat wavefront) is 82.8\%. We have not accounted for Fresnel loss at the face of the fiber and propagation loss through the fiber, which may explain part of the discrepancy. The remaining losses may be a result of uncalibrated higher order wavefront errors.

Next, we keep the DM map that optimizes injection into the SMF (which implies minimal wavefront aberration), but translate the injection stage to the location of the photonic lantern. We then tune the iris size to set the $F\#$ into the lantern, a procedure that we discuss below.

There are several metrics one could use to determine the optimal $F\#$. The signal in a given port is the sum of both the stellar throughput at the center ($\eta_s$) and the off-axis throughput at the planet location ($\eta_p$). Broadly, we wish to maximize the signal-to-noise ratio (i.e., the planet light relative to photon noise from stellar leakage), which scales as $\eta_p/\sqrt{\eta_s}$. While $\eta_s$ changes as a function of $F\#$, wavefront control can be used to further improve the null --- without significantly impacting the $F\#$ that maximizes the off-axis throughput for each port. Although implementing wavefront control with a PLN is left for future work, because it would give us additional control over $\eta_s$, we choose to optimize iris size for planet throughput instead. We expect the exoplanet light to mostly couple into one of the LP 11 ports (since they tend to have higher throughput overall), so the peak off-axis throughput ($\eta_{p_{\text{peak}}}$) through either the LP11a or the LP11b port is a simple proxy for exoplanet throughput. In our work, we maximize through the LP 11a mode (though the LP 11b mode would be an equally valid choice), resulting in a $F\#$ of 6.2.

In theory, the truly optimal $F\#$ depends on the planet location, and what the right metric is depends on how well that location is known. In practice, the $F\#$ of an instrument will be fixed to a certain value regardless of the target being observed, and most reasonable optimization metrics targeting close-in planets will result in similar $F\#$'s.

\subsection{Results} \label{sec:pln_results}

\subsubsection{Monochromatic}

In Figure 7a, we present the PLN throughput maps measured with 1568.772 nm light from the TLX2 tunable narrow linewidth laser, injected into the PoRT testbed with a PM fiber. The peak off-axis throughput of each port is reported in Table \ref{tab:port_summary_mono}.

We also simulate throughput maps based on the mode profiles we reconstructed in Section \ref{sec:oah} (note that the OAH data was taken using the same wavelength as the monochromatic PoRT measurements, but that the relative polarization between the OAH measurements and the PoRT measurements is unknown). We plot the simulated maps in Fig. 7b on the same color-scale as the PoRT measurements. We use a manual image alignment procedure to set the field of view, sampling, and rotation angle of the simulations to achieve the best match (across all six-ports) to the measured throughput maps.

The simulation with OAH modes assumes that the lantern is flux-preserving --- that whatever light gets coupled into a given port is maintained through the lantern. The real lantern assembly is lossy, so has lower throughputs than in simulation. Qualitatively, however, the simulated throughput maps and the measured throughput maps have similar morphologies, showing that our measurements of the PLN's behavior largely agree with the model.

To better resolve the central null, we repeat the scans, but with finer spatial sampling by a factor of 5. The results are shown in Fig. 7c. Because the maps are asymmetric due to manufacturing imperfections, and the apparent centers of the modes slightly offset from each other, it is ambiguous which XY coordinate, even in simulation, should be designated as the theoretical axial center to which the star would be aligned. One approach, which we use for this work, is to sum up the central throughput maps of the four nulled ports (shown in Fig. 7d), and then take the location of minimum summed throughput as the center. The stellar leakages ($\eta_s$) at the identified center for each port are reported in Table \ref{tab:port_summary_mono}.

In Figure 8, we plot select cross-sections of the throughput maps shown in Figure 7, and compare them against the throughput maps of an ideal MSPL with perfect LP modes. We show that the imperfect mode shapes cause different throughput profiles from an ideal MSPL, but that the measured profiles agree with the simulations using the modes obtained from OAH, except with additional losses from propagating through the lantern. Note that, while we do not see this phenomenon in our measured throughput maps, it is possible for the planet throughput at a given spatial location to be higher in one particular port than predicted with perfect LP modes. This is simply a result of the imperfect mode shapes, which distribute planet light differently amongst the ports, i.e., for a given planet position, the lantern imperfections can cause higher throughput in one port at the expense of throughput in other ports, relative to perfect LP modes.

In Table \ref{tab:port_summary_mono}, we also report the values of $\eta_s/\eta_{p_{\text{peak}}}$. Because the overall loss of each port cancels out in this ratio, it allows for a direct comparison between the PoRT measurements and the simulations using OAH modes. The quantity $\eta_s/\eta_{p_{\text{peak}}}$ is often referred to as the `null-depth' in discrete-aperture nulling interferometry contexts, though it is also called the `raw contrast' in fiber-fed spectroscopy contexts (which is different from how raw contrast is typically used in coronagraphy contexts). To avoid confusion over terminology, we will refer to it in this work explicitly as $\eta_s/\eta_{p_{\text{peak}}}$.

\begin{figure*}[t]
\begin{center}
    \includegraphics[scale=0.5]{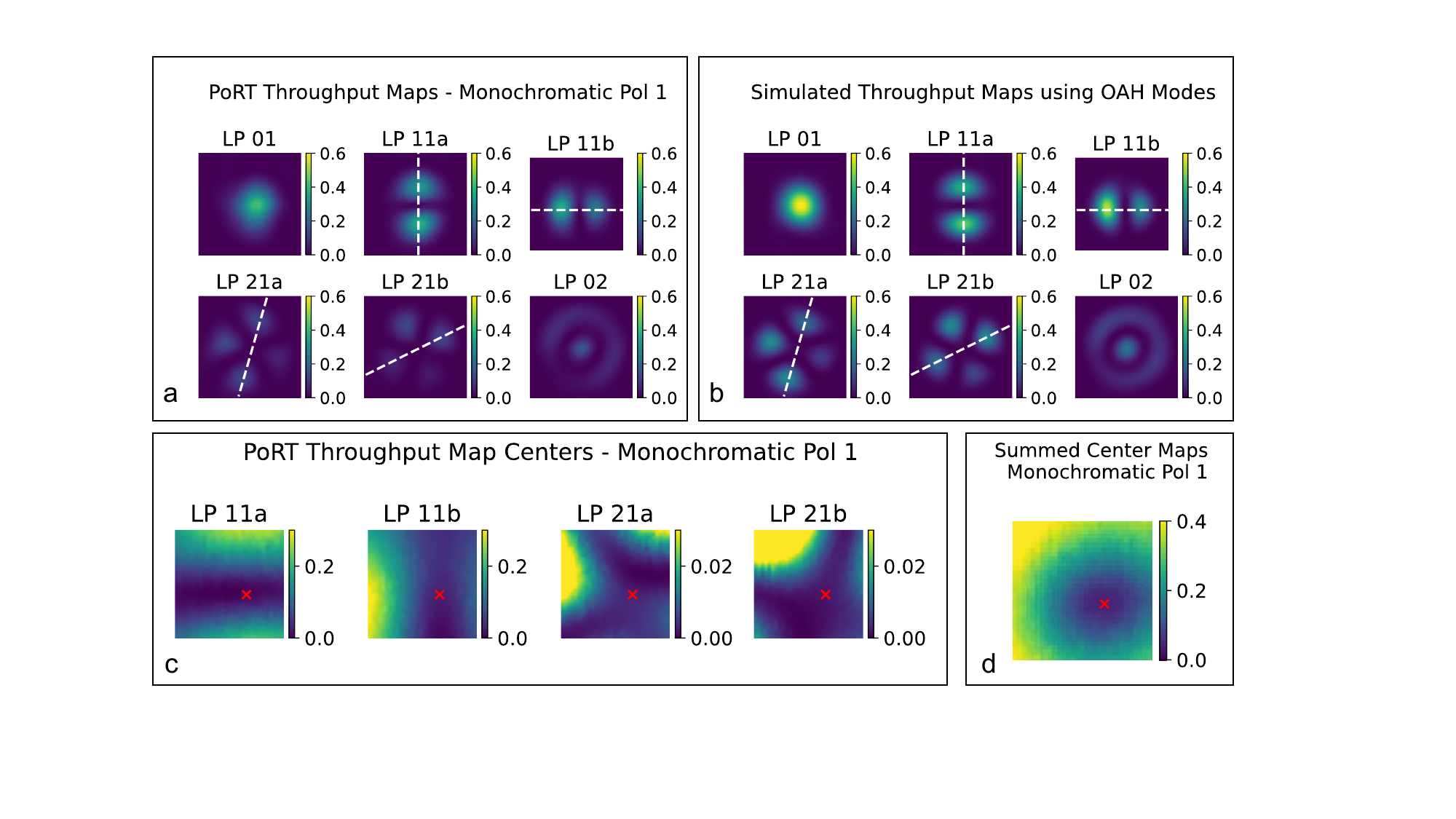}
	\caption{\label{fig:port_coupling_maps} a) Monochromatic PLN throughput maps measured with 1568.772 nm light from the TLX2 tunable narrow linewidth laser injected into the PoRT testbed with a PM fiber. White dashed lines indicate cross-sections plotted in Figure 8. b) Simulated throughput maps based on the mode profiles reconstructed using OAH at the same wavelength, assuming that the lantern is flux-preserving. White dashed lines indicate cross-sections plotted in Figure 8. c) Monochromatic PoRT throughput maps of the nulled ports with fine spatial sampling of the center. The red crosses indicate the axial center of the lantern, identified using the map in part (d). d) The summed throughput of the four maps in part (c). The location of minimum summed throughput is taken to be the lantern center, where $\eta_s$ is measured.
	}
\end{center}
\end{figure*}

\begin{figure*}[t]
\begin{center}
    \includegraphics[scale=0.6]{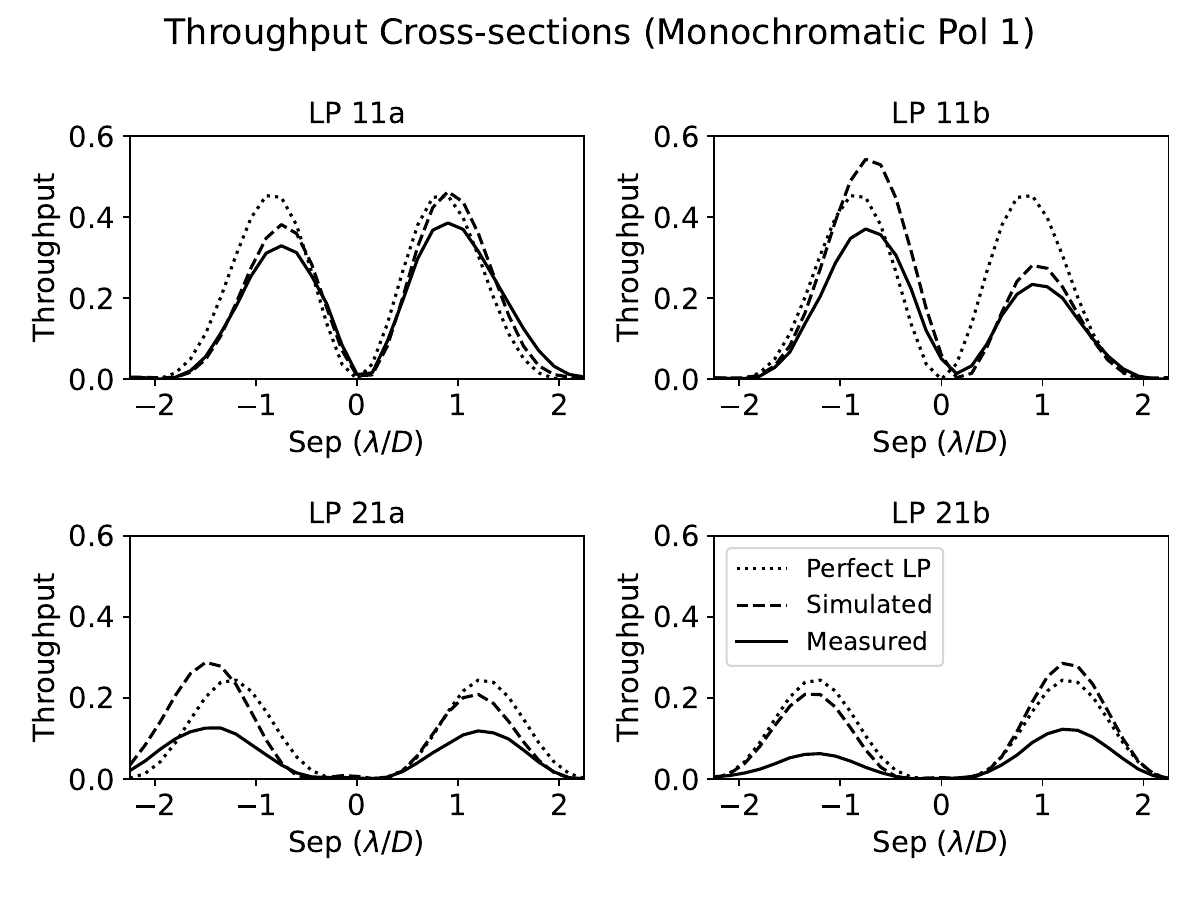}
	\caption{\label{fig:coup_cross_sections} Simulated and measured throughput cross-sections (as indicated in Figure 7) for the nulled ports of the PLN. For comparison, the ideal throughput cross-sections with perfect LP modes is also plotted. The measured PLN throughput diverges from the ideal PLN because of imperfect mode shapes. However, considering the overall throughput losses through the lantern reported in Table \ref{tab:throughputs}, the shapes of the measured throughput maps agree well with the predictions simulated using the measured modes. The shown measurements are with the monochromatic laser in the first polarization state, but the other measured profiles lie closely to the same curve.
	}
\end{center}
\end{figure*}

We then repeat the experiment using orthogonally polarized light by injecting the laser into the bench with a $90^{\circ}$ polarization rotating fiber (see Appendix \ref{app:thpt_maps} for the throughput maps, as well as a comparison between the maps obtained for the two polarizations). A summary of the important measurements is presented in Table \ref{tab:port_summary_mono}.

\begin{table}
\caption{Key monochromatic PLN metrics from a) simulations using the modes measured with OAH assuming that the lantern is flux-preserving, b) the PoRT testbed with the laser routed through a PM fiber (polarization 1), and c) the PoRT testbed with the laser routed through a $90^{\circ}$ polarization rotating fiber (polarization 2). $\eta_{p_{\text{peak}}}$ refers to the maximum throughput of each map, or what the throughput of a planet at the location of maximum coupling would be. For the nulled ports, $\eta_s$ refers to the throughput at the coaxial center of the lantern, corresponding to the stellar leakage. For the nulled ports, we also report the ratio $\eta_s/\eta_{p_{\text{peak}}}$, which is unaffected by the overall throughput of a port, and thus allows for a direct comparison between the PoRT measurements and the simulations using OAH modes.}

\begin{center} \label{tab:port_summary_mono}

\begin{tabular}{|c|c|c|c|c|c|c|} \hline 

 & \textbf{LP 01} & \textbf{LP 11a} & \textbf{LP 11b} & \textbf{LP 21a} & \textbf{LP 21b} & \textbf{LP 02} \\ \hline 

$\eta_{p_{\text{peak}}}$   (Sim.)& 0.610 & 0.463 & 0.542 & 0.285 & 0.288 & 0.232\\ \hline 

$\eta_{p_{\text{peak}}}$  (Pol. 1)& 0.421 & 0.385 & 0.384 & 0.138 & 0.130 & 0.160\\ \hline 

$\eta_{p_{\text{peak}}}$  (Pol. 2)& 0.461 & 0.396 & 0.379 & 0.146 & 0.140 & 0.171\\ \hline 

$\eta_s$  (Sim.)& N/A & $7.32 \times 10^{-4}$ & $3.65 \times 10^{-3}$ & $5.91 \times 10^{-3}$ & $9.05 \times 10^{-4}$ & N/A\\ \hline 

$\eta_s$  (Pol. 1)& N/A & $4.05 \times 10^{-3}$ & $2.64 \times 10^{-2}$ & $2.85 \times 10^{-3}$ & $1.77 \times 10^{-3}$ & N/A\\ \hline 

$\eta_s$  (Pol. 2)& N/A & $6.53 \times 10^{-3}$ & $3.00 \times 10^{-2}$ & $5.12 \times 10^{-4}$ & $1.07 \times 10^{-3}$ & N/A\\ \hline 

$\eta_s/\eta_{p_{\text{peak}}}$  (Sim.)& N/A & $1.59 \times 10^{-3}$ & $6.69 \times 10^{-3}$ & $2.05 \times 10^{-2}$ & $3.18 \times 10^{-3}$ & N/A\\ \hline 

$\eta_s/\eta_{p_{\text{peak}}}$  (Pol. 1)& N/A & $1.05 \times 10^{-2}$ & $6.88 \times 10^{-2}$ & $2.07 \times 10^{-2}$ & $1.37 \times 10^{-2}$ & N/A\\ \hline

$\eta_s/\eta_{p_{\text{peak}}}$  (Pol. 2)& N/A & $1.69 \times 10^{-2}$ & $7.82 \times 10^{-2}$ & $3.72 \times 10^{-3}$ & $8.24 \times 10^{-3}$ & N/A\\ \hline

\end{tabular}
\end{center}
\end{table}

\subsubsection{Broadband}
We also repeat the same experiment using a broadband Super-Luminescent Diode (SLD) light source with wavelength coverage from 1450 nm to 1625 nm (Thorlabs S5FC1550P-A2). The throughput maps for both polarizations are presented in the Appendix, along with a comparison between them. The peak and coaxial throughput values, as well as the ratio, determined using the same methodology as with monochromatic light, are reported in Table \ref{tab:port_summary_bb}. The broadband null-depths are not significantly different from the monochromatic ones (with a difference varying from a few percent to a factor of 2), indicating that the nulls are fairly achromatic. This is as expected: the MSPL is designed to be mode-selective across a wide wavelength range, and thus the (vortex-free) PLN is intrinsically achromatic, one of its major advantages compared to other nulling techniques.

\begin{table}
\caption{Key broadband PLN metrics from a) the PoRT testbed with the SLD routed through a PM fiber (polarization 1), and b) the PoRT testbed with the SLD routed through a $90^{\circ}$ polarization rotating fiber (polarization 2). $\eta_{p_{\text{peak}}}$ refers to the maximum throughput of each map, or what the throughput of a planet at the location of maximum coupling would be. For the nulled ports, $\eta_s$ refers to the throughput at the coaxial center of the lantern, corresponding to the stellar leakage. For the nulled ports, we also report the ratio $\eta_s/\eta_{p_{\text{peak}}}$.}

\begin{center} \label{tab:port_summary_bb}

\begin{tabular}{|c|c|c|c|c|c|c|}
\hline
 & \textbf{LP 01} & \textbf{LP 11a} & \textbf{LP 11b} & \textbf{LP 21a} & \textbf{LP 21b} & \textbf{LP 02} \\
\hline
$\eta_{p_{\text{peak}}}$  ( Pol. 1)& 0.438 & 0.404 & 0.374 & 0.132 & 0.124 & 0.177\\
\hline
$\eta_{p_{\text{peak}}}$  (Pol. 2)& 0.451 & 0.387 & 0.386 & 0.133 & 0.129 & 0.205\\
\hline
$\eta_s$  (Pol. 1)& N/A & $7.98 \times 10^{-3}$ & $3.46 \times 10^{-2}$ & $2.55 \times 10^{-3}$ & $1.26 \times 10^{-3}$ & N/A\\
\hline
$\eta_s$  (Pol. 2)& N/A & $8.54 \times 10^{-3}$ & $3.24 \times 10^{-2}$ & $2.52 \times 10^{-3}$ & $1.34 \times 10^{-3}$ & N/A\\
\hline
$\eta_s/\eta_{p_{\text{peak}}}$  (Pol. 1)& N/A & $2.06 \times 10^{-2}$ & $8.97 \times 10^{-2}$ & $1.92 \times 10^{-2}$ & $9.79 \times 10^{-3}$ & N/A\\
\hline

$\eta_s/\eta_{p_{\text{peak}}}$  (Pol. 2)& N/A & $2.21 \times 10^{-2}$ & $8.40 \times 10^{-2}$ & $1.89 \times 10^{-2}$ & $1.04 \times 10^{-2}$ & N/A\\

\hline
\end{tabular}
\end{center}
\end{table}

We did not perform OAH using the SLD, as the bandwidth of light results in a much smaller coherence length ($\sim 10$ microns), making matching path lengths practically infeasible.

\section{Discussion and Future Work}

In Table \ref{tab:port_summary_mono}, we also include the values for $\eta_{p_{\text{peak}}}$, $\eta_s$, and $\eta_s/\eta_{p_{\text{peak}}}$ obtained from the monochromatic simulation using OAH modes. The simulated value of $\eta_s/\eta_{p_{\text{peak}}}$ assumes no wavefront error, so is indicative of the limit imposed by the modal impurities of the lantern. For the nulled port measurements, $\eta_s/\eta_{p_{\text{peak}}}$ is generally higher on the testbed than in simulation. The difference suggests that the wavefront of the PoRT testbed, optimized for injection through a single-mode fiber, is still not perfectly flat. This is likely due to a combination of limited calibration precision, uncalibrated higher order wavefront modes, and testbed drifts between the time of SMF calibration and the time the PLN measurements were taken. However, a perfectly flat wavefront is not necessarily the optimal wavefront for a real PLN with modal impurities, as the deformable mirror can be used to partially compensate for the modal impurities of the lantern, and improve the null even beyond that predicted by a flat wavefront. For example, with monochromatic light in the second polarization, we measure an $\eta_s$ that is an order of magnitude than predicted by simulation. This suggests that the wavefront error in the system is interfering with the lantern mode in a way that deepens the stellar null. This also suggests that the difference in mode shape between the two polarizations is causing a difference in $\eta_s$ on the order of $\sim 10^{-3}$. Exploring using wavefront sensing and control schemes with the PLN, including in the presence of polarization differences, is left for future work. Although the MSPL does not lend itself to linear wavefront control \cite{lin_2022_plwfs}, it is a good candidate for data driven control such as Implicit Electric Field Conjugation \cite{haffert_2023_iefc} because of the low dimensionality from its limited number of ports.

The $\eta_s/\eta_{p_{\text{peak}}}$ values of $\sim 10^{-2}$ currently achieved by the PLN are approximately the same as the null-depth the VFN achieves on sky, limited by the adaptive optics residuals of the Keck II telescope \cite{Echeverri_KPICVFNComm}. Under these conditions, the VFN was able to tentatively detect a companion with flux ratio of 1/400 \cite{echeverri_2024}. Given that the PLN is expected to have even higher throughput, it could be used --- as is --- on ground-based telescopes, with signal-to-noise expected to exceed that of the VFN's. Additionally, the PLN has the potential to partially constrain the planet's location, a capability the VFN lacks \cite{xin_2022}, as well as the potential for image reconstruction similar to what has been explored for a non-mode-selective photonic lantern in Ref. \citenum{kim_2024}. Thus, future work includes testing and validating the PLN on-sky, as it can already be used to obtain scientifically useful observations of close-in giant exoplanets.

Like the VFN, the PLN's null-depths on-sky at Keck II would be expected to be limited by the residuals of the AO system. However, given a better adaptive optics system (or a space environment) the modal impurity or `cross-talk' of the lantern will become the dominant limiting factor. Wavefront control with one DM can only partially compensate for the cross-talk, as it would only modulate phase and not amplitude in the pupil plane. Wavefront control with two DMs is still chromatic, and thus expected to achieve nulls with smaller bandwidths than nulls resulting purely from the lantern's mode-selectivity. Therefore, more work should be invested in creating MSPLs with lower levels of cross-talk, as this would result in deeper, naturally broadband nulls that would relieve additional demand on the wavefront control system \cite{astrophotonics_roadmap}.

Lastly, although the lantern we use in this work is optimized for 1550 nm, silica-based mode-selective lanterns can be designed to operate at different wavelengths, ranging from the visible spectrum up to about 2 um. However, other fiber technologies \cite{price_nonsilica} would be needed to access wavelengths outside of this range.

\section{Conclusion}

In this work, we characterize the properties of a MSPL optimized around 1550nm, and perform the first laboratory demonstration of a PLN. We measure the throughput maps for each port (in two polarizations for both monochromatic and broadband light) and calculate the null-depths of the nulled ports, which are around the $10^{-2}$ level. We find that the mode shapes measured using off-axis holography can be used in simulations to model and predict the behavior of a real PLN. Future work involves using wavefront sensing and control to further improve the null-depths, as well efforts towards improving the modal purity of the lanterns themselves. In the meantime, the photonic lantern nuller already achieves broadband nulls suitable for observing young gas giants at the diffraction limit using ground-based observatories, a capability that should be tested on-sky.

\appendix

\section{Additional Throughput Maps and Polarization Comparison} \label{app:thpt_maps}

We present the throughput maps measured using monochromatic light in the second polarization state (with the $90^\circ$ polarization rotating fiber) in Fig. 9, using broadband light in the first polarization state (with the PM fiber) in Fig. 10, and using broadband light in the second polarization state in Fig. 11.

\begin{figure*}[t]
\begin{center}
    \includegraphics[scale=0.48]{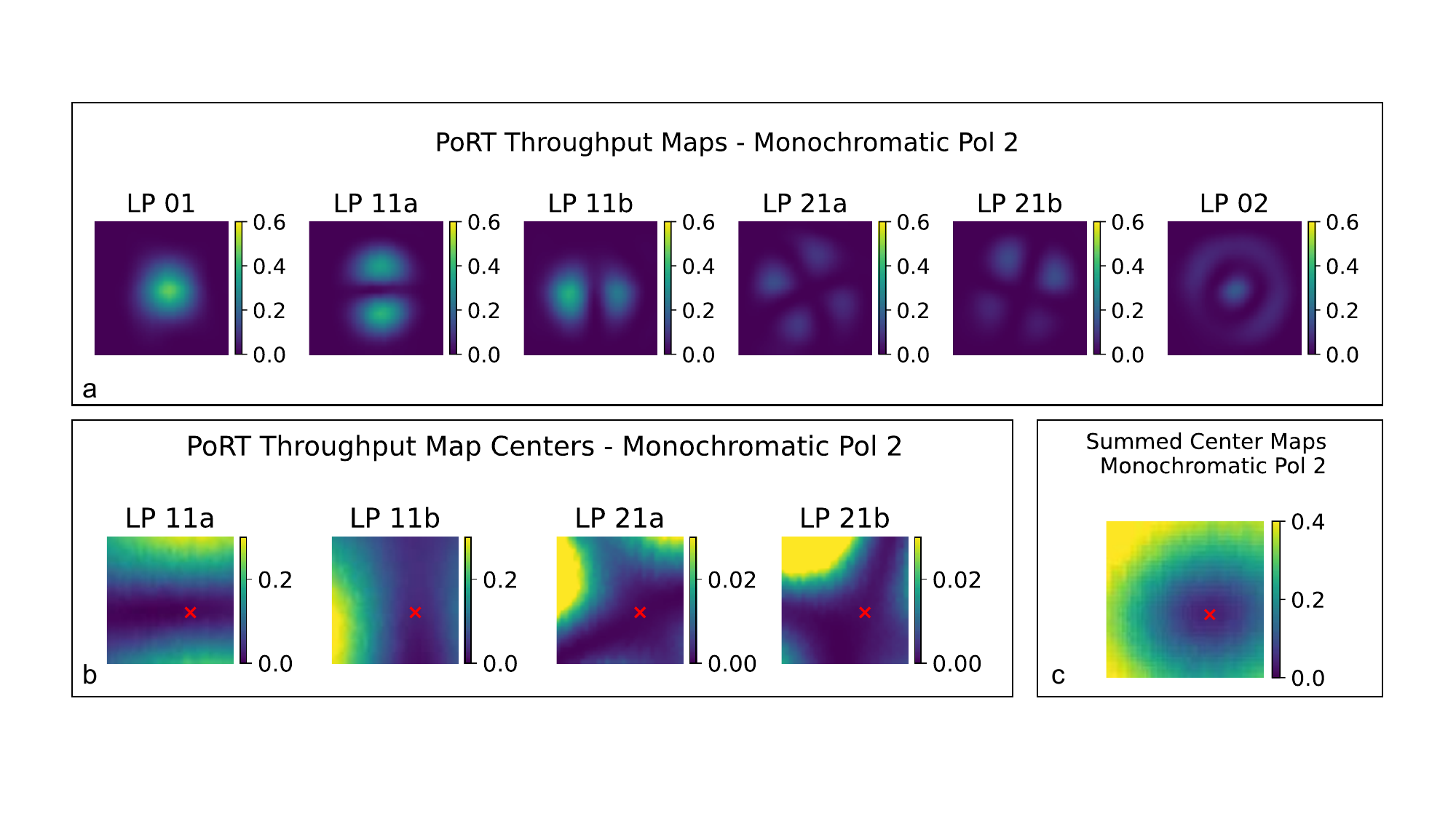}
	\caption{\label{fig:port_coupling_maps_pol2} Monochromatic PLN throughput maps measured with 1568.772 nm light from the TLX2 tunable narrow linewidth laser injected into the PoRT testbed with a $90^{\circ}$ polarization rotating fiber, such that the polarization is orthogonal to that of Fig. 7. a) Throughput maps of all ports across the PLN field of view. b) Throughput maps of the nulled ports with fine spatial sampling of the center (note that the LP 21 maps are on a different color scale from the LP 11 maps). The red crosses indicate the axial center of the lantern, identified using the map in part (c). c) The summed throughput of the four maps in part (b). The location of minimum summed throughput is taken to be the lantern center, where $\eta_s$ is measured.
	}
\end{center}
\end{figure*}

\begin{figure*}[t]
\begin{center}
    \includegraphics[scale=0.48]{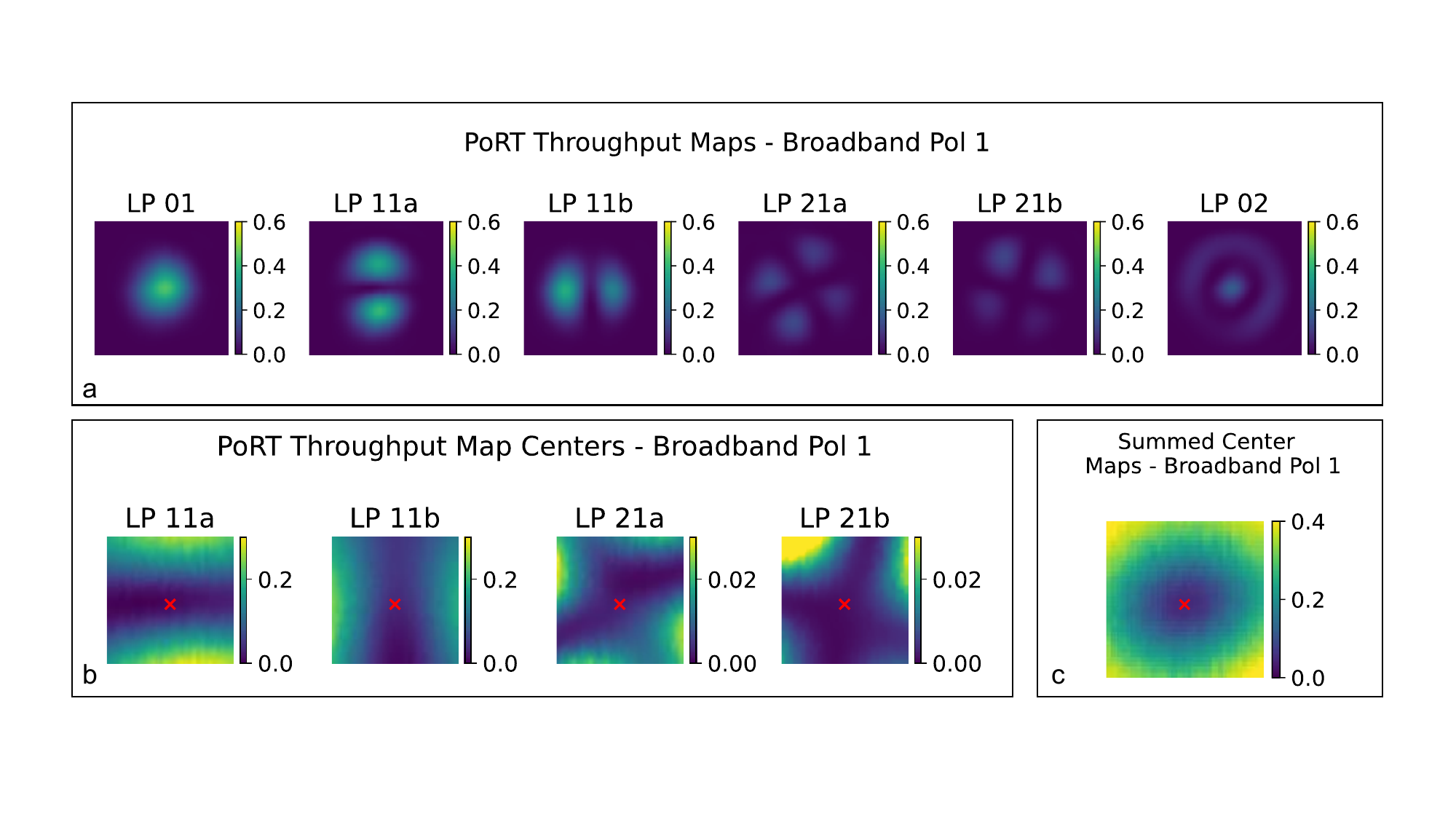}
	\caption{\label{fig:port_coupling_maps_broadband_pol1} Laboratory PLN results measured on PoRT, using an SLD light source from 1450 nm to 1625 nm injected with a PM fiber. a) Throughput maps of all ports across the PLN field of view. b) Throughput maps of the nulled ports with fine spatial sampling of the center (note that the LP 21 maps are on a different color scale from the LP 11 maps). The red crosses indicate the axial center of the lantern, identified using the map in part (c). c) The summed throughput of the four maps in part (b). The location of minimum summed throughput is taken to be the lantern center, where $\eta_s$ is measured.
	}
\end{center}
\end{figure*}

\begin{figure*}[t]
\begin{center}
    \includegraphics[scale=0.48]{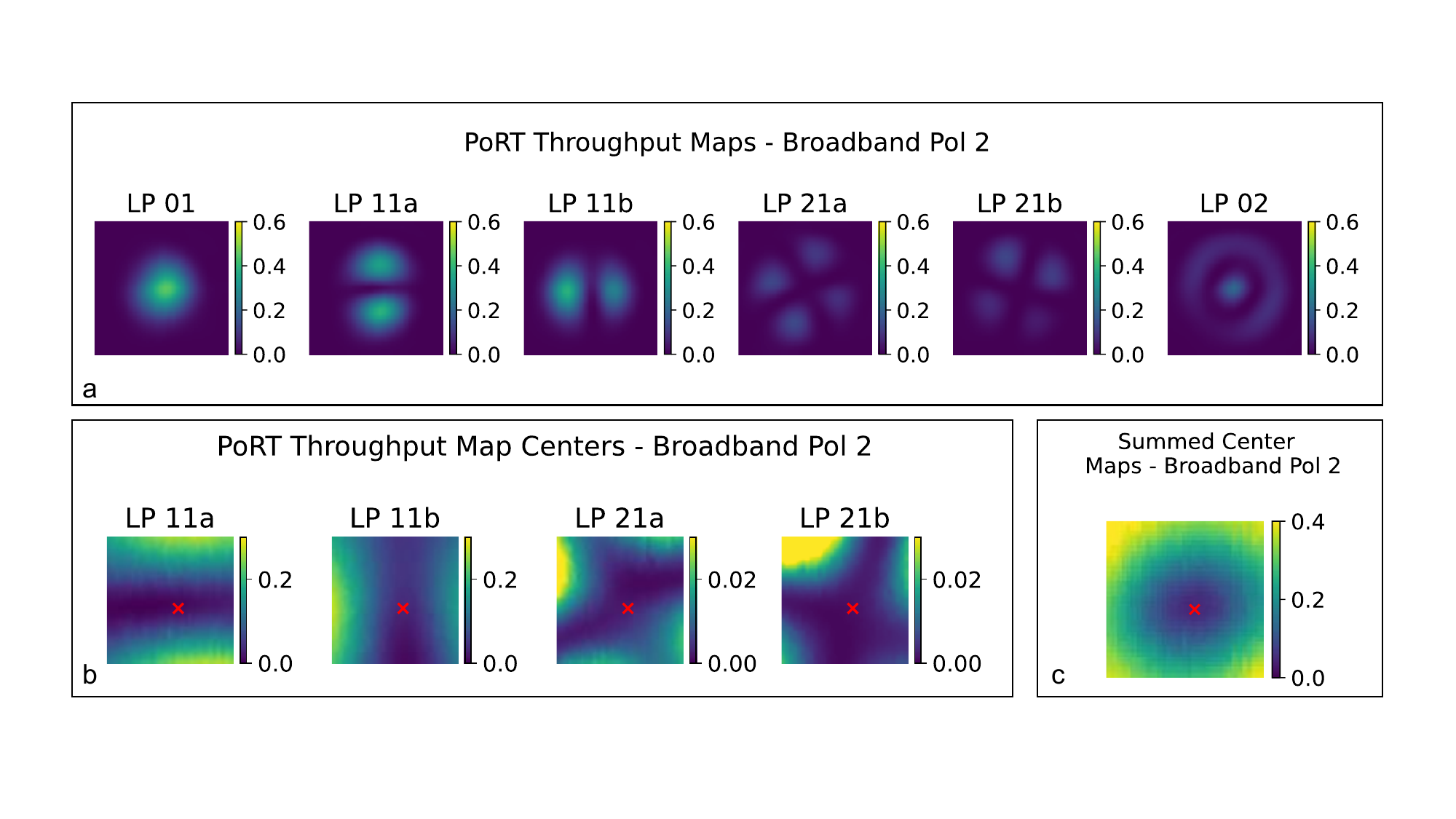}
	\caption{\label{fig:port_coupling_maps_broadband_pol2} Laboratory PLN results measured on PoRT, using an SLD light source from 1450 nm to 1625 nm injected with a $90^{\circ}$ polarization rotating fiber. a) Throughput maps of all ports across the PLN field of view. b) Throughput maps of the nulled ports with fine spatial sampling of the center (note that the LP 21 maps are on a different color scale from the LP 11 maps). The red crosses indicate the axial center of the lantern, identified using the map in part (c). c) The summed throughput of the four maps in part (b). The location of minimum summed throughput is taken to be the lantern center, where $\eta_s$ is measured.
	}
\end{center}
\end{figure*}

In Fig. 12, we plot the difference in the finely-sampled central throughput maps between the two orthogonal polarization states (the maps for polarization 1 are subtracted from the maps for polarization 2).

\begin{figure*}[t]
\begin{center}
    \includegraphics[scale=0.5]{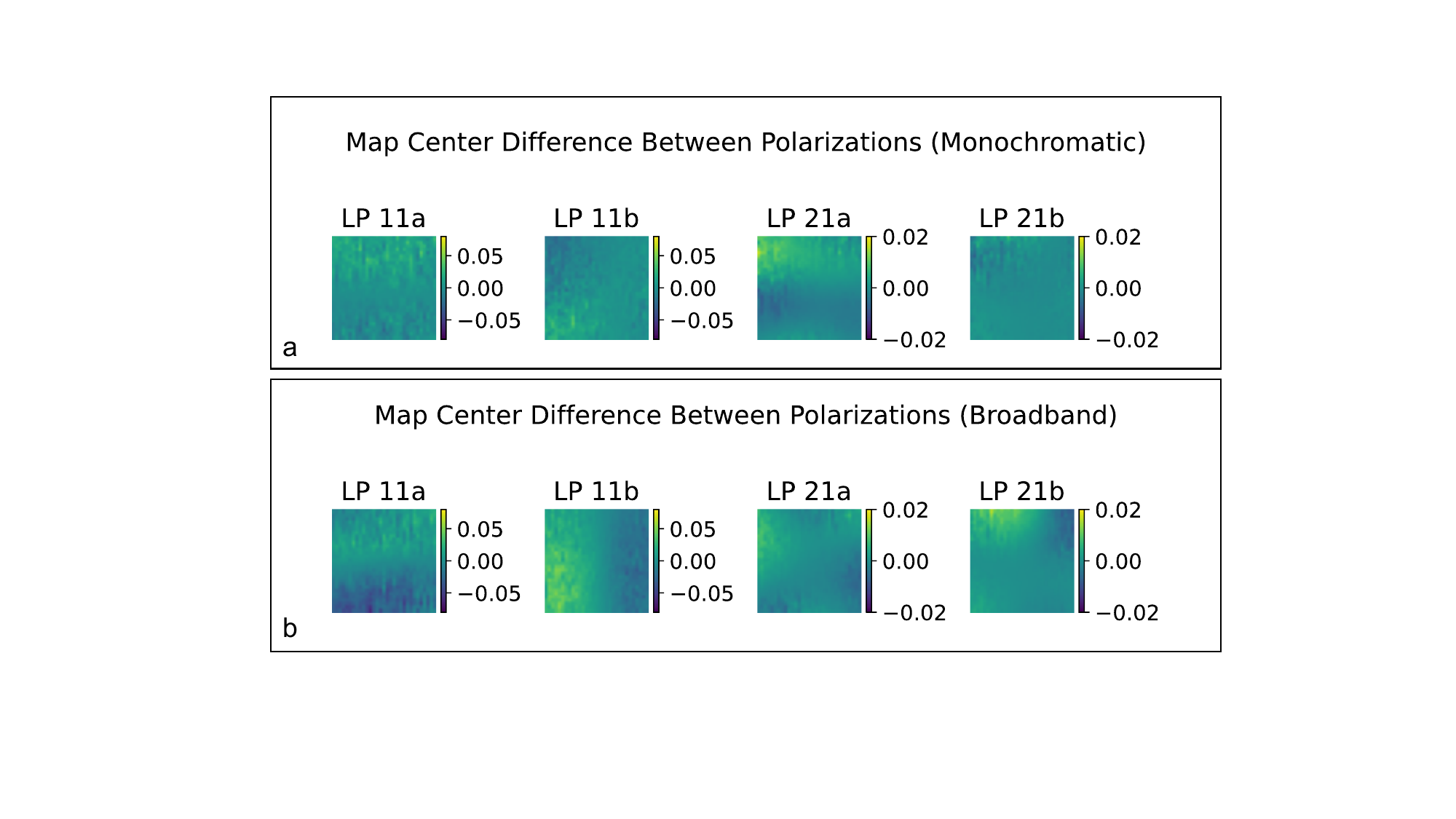}
	\caption{\label{fig:pol_diff} Difference in the finely-sampled central throughput maps between two orthogonal polarization states, with throughput maps for polarization 1 subtracted from the maps for polarization 2, using a) monochromatic light and b) broadband light. The monochromatic null-depth of the LP 21a port with the second polarization state is almost an order of magnitude deeper than the first, suggesting that the lantern mode shapes do depend on the polarization at some level. The limitations that these polarized differences impose on the null that wavefront sensing and control can achieve should be examined as part of future work.
	}
\end{center}
\end{figure*}

When observing the key measurements in Table \ref{tab:port_summary_mono} and \ref{tab:port_summary_bb}, we find that the differences between the two polarization states are slight for most of the ports. However, the monochromatic null-depth of the LP 21a port shows a stark difference, with the second polarization state exhibiting a null-depth that is almost an order of magnitude deeper than the first. This result suggests that the lantern mode shapes might depend on the polarization at some level. The limitations that these polarized differences impose on the null that wavefront sensing and control can achieve should be examined as part of future work.

One caveat is that the throughput maps presented were measured sequentially, not simultaneously. However, the measurements for polarization state 1 and for polarization state 2 were obtained about several hours apart, over which the testbench is relatively stable. We confirmed this stability by measuring the finely sampled LP21b map in both polarization states, one immediately after the other, and confirming that the difference map is consistent (in both features and magnitude) with the difference map obtained using the original data taken hours apart.

% \disclosures 
\subsection*{Disclosures}
The authors have no relevant financial interests in the manuscript and no other potential conflicts of interest to disclose.

\subsection* {Acknowledgments}    
 
Y.X. acknowledges support from the National Science Foundation Graduate Research Fellowship under Grant No. 1122374. Additional effort has been supported by the National Science Foundation under Grant Nos. 2109231 and 2308360. This research was carried out in part at the California Institute of Technology and the Jet Propulsion Laboratory under a contract with the National Aeronautics and Space Administration (NASA). This research made use of hcipy \cite{por2018hcipy}; Astropy \cite{astropy:2013,astropy:2018,astropy:2022}; NumPy \cite{harris2020array}; SciPy \cite{2020SciPy-NMeth}; and Matplotlib \cite{Hunter:2007_matplotlib}.

\subsection* {Code, Data, and Materials Availability}
The code and data used in this work can be found at \url{https://github.com/yinzi-xin/pln_lab_jatis}, or with the DOI: 10.5281/zenodo.10888625.

%%%%% References %%%%%

\bibliography{report}   % bibliography data in report.bib
\bibliographystyle{spiejour}   % makes bibtex use spiejour.bst

%%%%% Biographies of authors %%%%%

\vspace{2ex}\noindent\textbf{Yinzi Xin} is a graduate student at Caltech, where she works in the Exoplanet Technology Lab under the guidance of her advisor, Dimitri Mawet. Her research interests lie in the field of high contrast imaging instrumentation and data analysis for exoplanets. She is interested in wavefront sensing and control, coronagraphy, and the development of new instrument concepts. 

\vspace{1ex}
\noindent Biographies and photographs of the other authors are not available.

\end{spacing}
\end{document}